\documentclass{nature}
\usepackage{graphicx}
\usepackage{dcolumn}
\usepackage{bm}
\usepackage{physics}
\usepackage{amsmath}
\usepackage{amssymb}
\usepackage{array}
\usepackage{xcolor}
\usepackage{times}
\newcolumntype{P}[1]{>{\centering\arraybackslash}p{#1}}
\usepackage[colorlinks=true, breaklinks=true, linkcolor=blue, citecolor=blue, urlcolor=blue]{hyperref}
\usepackage[normalem]{ulem}
\usepackage{lineno}

\usepackage{graphicx}
\makeatletter
\let\saved@includegraphics\includegraphics
\AtBeginDocument{\let\includegraphics\saved@includegraphics}
\renewenvironment*{figure}{\@float{figure}}{\end@float}
\makeatother

\newcommand{\bonnpi}{Physikalisches Institut, University of Bonn, Nussallee 12, 53115 Bonn, Germany}
\newcommand{\geneva}{Department of Quantum Matter Physics, University of Geneva, Quai Ernest-Ansermet 24, 1211 Geneva, Switzerland}
\newcommand{\unbc}{Department of Physics, University of Northern British Columbia, Prince George, BC V2N 4Z9 Canada}
\newcommand{\dortmund}{Department of Physics, TU Dortmund University, 44227 Dortmund, Germany}
\newcommand{\koln}{Institute of Physics II, University of Cologne, 50937 Cologne, Germany}
\newcommand{\ausburg}{Experimental Physics V, Center for Electronic Correlations and Magnetism, Institute of Physics, University of Augsburg, 86135 Augsburg, Germany}
\newcommand{\dresdenMFL}{Dresden High Magnetic Field Laboratory (HLD-EMFL), Helmholtz-Zentrum Dresden-Rossendorf (HZDR), 01328 Dresden, Germany}
\newcommand{\dresdenIRP}{Institute of Radiation Physics, Helmholtz-Zentrum Dresden-Rossendorf (HZDR), 01328 Dresden, Germany}

\title{Experimental observation of repulsively bound magnons}

\begin{document}

\maketitle

\author{Zhe Wang$^{1, 2, 3, 9}$, Catalin-Mihai Halati$^{4, 5}$, Jean-S\'ebastien Bernier$^{5, 6}$, Alexey Ponomaryov$^{7, 8}$, Denis I. Gorbunov$^{7}$,
  Sandra Niesen$^{2}$, Oliver Breunig$^{2}$, J. Michael Klopf$^{8}$, Sergei Zvyagin$^{7}$, Thomas~Lorenz$^{2}$, Alois~Loidl$^{3}$, Corinna Kollath$^{5}$}

\begin{affiliations}
 \item \dortmund 
 \item \koln 
 \item \ausburg 
 \item \geneva 
 \item \bonnpi 
 \item \unbc 
 \item \dresdenMFL 
 \item \dresdenIRP 
 \item e-mail: zhe.wang@tu-dortmund.de
\end{affiliations}


\clearpage

\setlength{\parskip}{4pt}



\begin{abstract}
  Stable composite objects (e.g. hadrons, nuclei, atoms, molecules, and superconducting pairs) formed by attractive forces are ubiquitous in nature. 
  In contrast, composite objects stabilized via repulsive forces  were long thought to be theoretical constructions due to their fragility in naturally occurring systems. 
  Surprisingly, the formation of bound atom pairs by strong
  {\it repulsive} interactions has been demonstrated experimentally in optical lattices\cite{WinklerZoller2006}.
  Despite this success, repulsively bound particle pairs were believed to have no analogue in condensed matter due to strong decay channels.
  Here, we present spectroscopic signatures of repulsively bound three-magnon states and bound magnon pairs, in the Ising-like chain antiferromagnet BaCo\textsubscript{2}V\textsubscript{2}O\textsubscript{8}.
  In large transverse fields, below the quantum
  critical point, we identify repulsively bound magnon states by comparing terahertz spectroscopy measurements
  to theoretical results for the Heisenberg-Ising chain antiferromagnet,
  a paradigmatic quantum many-body model\cite{Pfeuty1970, SachdevBook2011, DuttaSen2015, MussardoBook2020}.
  Our experimental results show that these high-energy repulsively bound magnon states are
  well separated from continua, exhibit significant dynamical responses and, despite dissipation, are sufficiently long-lived to be identified.
  As the transport properties in spin chains can be altered by magnon bound states, we envision 
such states could serve 
  as resources for magnonics based quantum information processing technologies\cite{Subrahmanyam2004,
  BarmanWinklhofer2021,YuanYan2022}. 
\end{abstract}

The stability of attractively bound pairs is typically rooted in their binding forces lowering the energy of the state.
For example, in ferromagnets, spin waves whose quantization corresponds to a bosonic quasiparticle, a magnon,
are the conventional excitations. 
There, due to the attractive ferromagnetic interaction, two magnon excitations can remain bound together forming a two-magnon attractively bound state\cite{Bethe1931, Wortis1963, Hanus1963}.
In contrast, the formation of stable composite objects due to repulsive interaction typically
{\it increases}  the system energy.
For cold bosonic atoms in optical lattices,
repulsively bound on-site pairs dressed with quantum fluctuations become stable eigentstates
when the on-site repulsive interaction between atoms is significantly larger than the tunneling rate.
As these atoms are only weakly coupled to dissipative channels, such bound
objects have been observed to move through the lattice as dressed pairs for relatively long times 
because their large repulsion energy cannot be converted into kinetic energy\cite{WinklerZoller2006, DeuchertCederbaum2012}.

Whereas many condensed matter systems host attractively bound pairs, the realization of repulsively bound states in solid-state materials, 
in the absence of additional symmetry protection, has been considered as impossible due to multiple relaxation channels. Here we overcome this challenge by choosing carefully the compound to optimize these states stability.
We report on experimental signatures of repulsively bound
two- and three-magnon states in the antiferromagnetic compound
BaCo\textsubscript{2}V\textsubscript{2}O\textsubscript{8} modeled, under a strong transverse magnetic field, as a Heisenberg-Ising spin-$1/2$ chain.

BaCo\textsubscript{2}V\textsubscript{2}O\textsubscript{8} is exceptionally suited
for the observation of repulsively bound states due to 
several properties\cite{NiesenLorenz2013,KimuraWatanabe2013,
WangLoidl2018b}:
(i) nearest-neighboring spins are antiferromagnetically coupled\cite{NiesenLorenz2013,KimuraWatanabe2013}, leading to a repulsive interaction between magnons;
(ii) the transverse-field induced transition to the polarized phase is experimentally within reach\cite{WangLoidl2018b}; 
(iii) due to the material crystallographic structure, the spins are subjected to an effective staggered magnetic
field in addition to the uniform component\cite{NiesenLorenz2013,KimuraWatanabe2013}; 
(iv) with increasing field, the one-dimensionality becomes more pronounced, effectively enhancing its strongly correlated nature\cite{WangLoidl2018b}.

This last point is particularly crucial as this compound is described by the paradigmatic transverse-field Heisenberg-Ising spin-$1/2$ chain model.
This model is of tremendous importance for understanding a variety of physical phenomena
including quantum critically and duality\cite{SachdevBook2011, DuttaSen2015,  MussardoBook2020},
topological excitations\cite{FaureGrenier2018,TakayoshiGiamarchi2018} and quench
dynamics\cite{CalabreseFagotti2011, Caux2016, JamesRobinson2019, TanMonroe2021}, and has profoundly impacted various branches of physics, ranging from condensed matter, cold atoms to quantum information.
Here we show that this model has yet to reveal some of its secrets, as we demonstrate that it can even host two- and three-magnon repulsively bound states.

A simplified sketch of two- and three-magnon bound states above the spin-polarized ground state (Fig.~\ref{fig:sketch_results}c) realized by the application
of a strong magnetic field is presented in Fig.~\ref{fig:sketch_results}e-f. Due to the presence of an antiferromagnetic
exchange coupling, the energy of two neighboring magnons (Fig.~\ref{fig:sketch_results}e) is {\it higher} than the energy of two unbound spatially
distant ones (Fig.~\ref{fig:sketch_results}d). In BaCo\textsubscript{2}V\textsubscript{2}O\textsubscript{8},
the energy difference between the bound and unbound states is further increased 
by the presence of a spatially dependent effective magnetic field, a crucial ingredient for their experimental observation.
Figure~\ref{fig:sketch_results}e-f only depict a simplified classical representation of magnons and
quantum bound states, neglecting their dressing by quantum fluctuations.

In Fig.~\ref{fig:sketch_results}g, the experimentally detected modes are represented by their frequencies versus applied transverse field. 
Comparing to a one-dimensional Heisenberg-Ising (or XXZ) model, we demonstrate that the observed higher energy modes marked by \textit{D} and \textit{T} are associated with two-
and three-magnon repulsively bound states, respectively. 
In the experimentally accessed low-temperature and high-field regime, the long-range magnetic order stabilized due to small interchain couplings has been suppressed\cite{WangLoidl2018b}.
Hence, we neglect small interchain couplings present in real materials, and describe each spin chain by the one-dimensional XXZ spin-1/2 model
\begin{equation}
\label{eq:Hamiltonian}
H = J\sum_{j=1}^{L-1} \left( S_j^x S^x_{j+1}+ S_j^y S^y_{j+1} +\Delta S_j^z S^z_{j+1}\right)
- \mu_B B_x \sum_{j=1}^L  (g_j^{xx} S^x_j+g_j^{xz} S^z_j ), 
\end{equation}
with an antiferromagnetic exchange, $J > 0$, and an Ising-like
anisotropy, $\Delta > 1$ (Ref.\cite{ShibaKindo2003,KimuraWatanabe2013, NiesenLorenz2013, NiesenLorenz2014}).
$S_j^{x,y,z}$ denote the spin components at the $j$-th site of the chain with length $L$.
The $x$-coordinate is defined parallel to the applied  
field $B_x$ along $[110]$, while $y$ and $z$ are
along the $[-110]$ and $c$ directions, respectively. 
The last term describes the Zeeman interaction, given the Landé  \textit{g}-factors $g^{xx}_j$ and $g^{xz}_j$ and the Bohr magneton $\mu_B$.
The anisotropic $g$-values result from the edge-sharing CoO\textsubscript{6} octahedra\cite{KimuraWatanabe2013, NiesenLorenz2013}, which form a four-fold screw chain along the $c$ direction (Fig.~\ref{fig:sketch_results}a).

Within this isolated chain model, the positive exchange $J$ induces an antiferromagnetic ground state at zero field (Fig.~\ref{fig:sketch_results}b). In this regime, the elementary excitations
are spinons\cite{FaddeevTakhtajan1981,
  TennantNagler1993, StoneTurnbull2003, MourigalRonnow2013, WuAronson2016} carrying a fractional quantum number of spin-$1/2$.
In the applied transverse field, a phase transition into a quantum paramagnetic phase occurs at a critical field $B_c$.
This state is characterized by a field-induced polarization of the spins (Fig.~\ref{fig:sketch_results}c).
According to measurements of magnetization, 
sound velocity and magnetocaloric effect\cite{WangLoidl2018b}, the quantum phase transition of a transverse-field Ising nature\cite{SachdevBook2011, DmitrievLangari2002}
occurs in BaCo\textsubscript{2}V\textsubscript{2}O\textsubscript{8} at $B_c = 40~$T.
On the polarized ground state, the low-energy excitations are magnons, and due to quantum fluctuations a finite
density of spin flips is already present in this ground state. 
Since the magnetization in the field direction
is important at large fields even below $B_c$, the excitations in the regime $0 \ll B < B_c$ can be described 
in terms of magnons\cite{HalatiBernier2023}.

In this work, we address the following fundamental question: Can repulsively bound states of two or three magnons be formed 
when the strongly fluctuating ground state is perturbed by a photon or neutron flipping an additional spin?
For strong transverse fields, as the effective interaction between magnons is repulsive, i.e. $\propto J>0$, the energy 
of such magnon bound states would be larger than that of unbound magnons. In principle, many such high energy states should exist. 
However, to be sufficiently stable and detectable, these states need to be well separated from excitation continua. 
This condition is fulfilled in BaCo\textsubscript{2}V\textsubscript{2}O\textsubscript{8} where staggered magnetic-field components 
are induced by the site-dependent $g$-factors, $g^{xx}_j = g^x_u-(-1)^j g^x_s$ and $g^{xz}_j=g^z \sin(\frac{\pi}{2}j)$ 
(see Methods).

\section*{Quantum spin dynamics around $B_c$}
We present below our experimental results highlighting the presence of repulsively bound magnon states in
BaCo\textsubscript{2}V\textsubscript{2}O\textsubscript{8}. This compound is excited with magnetic field of terahertz
photons ($1$~THz $\sim$ $4.1$~meV) in applied pulsed fields up to $61$~T. 
The obtained absorption spectra are displayed in Fig.~\ref{fig:exp_results_1}a.  
At $1.478$~THz, the spectrum exhibits one resonance peak (labeled $M^u_{\pi/2}$)
at $42.1$~T above $B_c$, and below $B_c$ two additional peaks at $29.8$~T and $22.8$~T, labeled as $M_0^u$ and $D_{\pi}$, respectively.
All three modes shift to higher fields with increasing frequencies. The $M^u_{\pi/2}$ mode is sharply resolved above $B_c$,
while at low fields it is hardly traceable. In contrast, the other two modes ($M_0^u$ and $D_{\pi}$)
can still be clearly detected at lower fields. At much lower frequencies, e.g.~for $0.120$~THz (Fig.~\ref{fig:exp_results_1}b),
our electron-spin-resonance spectroscopy reveals two resonance peaks at $30.1~$T and $43.4~$T, above and below $B_c$, respectively. By decreasing the frequency, we observe both peaks approaching the critical field, an indication of gap closure at $B_c$.

\section*{Quantum spin dynamics at $0 \ll B < B_c$}
We explore the quantum spin dynamics at even higher energies by carrying out magneto-optic
measurements in strong static magnetic fields up to $32~$T. As shown in Fig.~\ref{fig:exp_results_2}a, the
absorption spectrum at $2$~K and $32$~T is characterized by two peaks at $1.56~$THz and $1.90~$THz, precisely corresponding
to the resonance modes $M^u_0$ and $D_{\pi}$, revealed by the field-sweeping electron-spin-resonance technique (Fig.~\ref{fig:exp_results_1}a). 
The modes $M^u_0$ and $D_{\pi}$ soften as the field is decreased. In contrast, at $24~$T another lower-frequency mode $M^l_{\pi/2}$
appears, with substantially smaller spectral weight, which evolves towards higher frequencies at lower fields. 
In the highest available frequency range above $2~$THz (whose photon energy is $> 3J$), we resolve two additional modes in the spectra (Fig.~\ref{fig:exp_results_2}b-c). For example at $22~$T, the
mode $D_{\pi/2}$ is found at $2.47~$THz and the mode $T_{\pi/2}$ at $2.92~$THz. We can unambiguously resolve these
two modes by their systematic shifts in the applied fields, even though they are characterized by a relatively small spectral weight. 

\section*{Many-body numerical simulations}
The experimental resonance frequencies versus magnetic fields are extracted for all these 
excitations and summarized in Fig.~\ref{fig:sketch_results}g. 
We identify the nature of these modes by comparing them to our precise theoretical results (Fig.~\ref{fig:sketch_results}h).
The theoretical modes are obtained by computing the dynamical spin structure 
factor $\mathcal{S}(q,\omega)$ (see Methods for the definition)
describing the response to a single spin-flip triggered by a linear coupling to the terahertz magnetic field
and obeying the selection rule $\Delta S = \pm 1$. 
Due to the four-fold screw symmetry (Fig.~\ref{fig:sketch_results}a), 
the zero-momentum transfer in the reduced zone scheme, 
probed by terahertz spectroscopy, corresponds to quasi-momenta $0,\pi/2,\pi$ in the extended Brillouin zone. 
The dynamical structure factor is derived from
many-body numerical simulations of the quantum spin dynamics for the Hamiltonian
in Eq.~(\ref{eq:Hamiltonian}) using the numerically exact time-dependent matrix product state (tMPS) algorithm\cite{Daley2014, WhiteFeiguin2004, Schollwoeck2011}.
Our results are presented in Fig.~\ref{fig:theo_results}a-c for the three regimes $0 \ll B < B_c$, $B \approx B_c$ and $B > B_c$, respectively.
Corresponding to the intensity maxima (as marked by green circles at the quasi-momenta $q=0$, $\pi/2$, and $\pi$ in Fig.~\ref{fig:theo_results}), the frequencies extracted from the theoretical spectral function are shown 
in Fig.~\ref{fig:sketch_results}h for the comparison to the experimentally observed modes.

\section*{Comparison between experiment and theory}
Very good agreement is achieved for all seven detected modes, using the parameters $J=2.82~$meV, $\Delta=1.92$, $g^x_u = 3.06$, $g^x_s = 0.66$, and $g^z = 0.21$. These values also provide an excellent description for the field-dependent magnetization (Extended Data Fig.~\ref{fig:magnetization}) and are consistent with previously reported values\cite{KimuraItoh2008,
  CanevetLejay2013, KimuraWatanabe2013, NiesenLorenz2013, GrenierLejay2015, FaureGrenier2018,
  FaurePetit2019, WangLoidl2018b, WangLoidl2019, FaureSimonet2021, OkutaniHagiwara2021} (Extended Data Table 1 and Extended Data Table 2).
We find that well above the critical field ($B \gg B_c$), the spins are polarized
by the magnetic field (Fig.~\ref{fig:sketch_results}c). In this regime, the spin dynamics is characterized by unbound magnon excitations, see Fig.~\ref{fig:theo_results}c and the sketch in Fig.~\ref{fig:sketch_results}d. 
However, instead of a single band, we obtain two distinct excitation bands, as denoted by $M^l$ and $M^u$ in Fig.~\ref{fig:theo_results}c.
The gap between the two bands is due to the site-dependent, staggered effective field, i.e. $(-1)^jg_s^xB_x$ (Extended Data Fig.~\ref{fig:staggering_45T}).
Both bands correspond to unbound magnons, excited by the terahertz field, propagating along the chain.

With decreasing field towards $B_c$, both magnon bands shift to lower energies as seen experimentally and theoretically.
The lower mode marked by $M_\pi^l$ decreases approximately linearly and becomes gapless at $B_c$ signaling the quantum phase transition.
The gap of this mode reopens more slowly below $B_c$. A linear extrapolation of the field dependence above the $B_c$ determines a critical field of $B_c = 40~$T (see Fig.~\ref{fig:sketch_results}g), in excellent agreement with thermodynamic measurements\cite{WangLoidl2018b}.

Close to the critical field, $B\approx B_c$, we identify new excitations (denoted by $D$ in Fig.~\ref{fig:theo_results}b) at energies higher than the unbound magnon bands.
The characteristic energies of the $D$ modes are neither simply twice that of the $M^l$ mode, nor of the $M^u$ mode. 
To analyze the nature of the excitations, we monitor theoretically the expectation distance between the
flipped spins present in the states contributing most to this dynamical-structure-factor feature\cite{HalatiBernier2023}.
As in a bound state the magnons should mostly be confined next to each other, the distance should be small\cite{HalatiBernier2023} (e.g.~$<2.5$ sites, see Fig.~\ref{fig:sketch_results}e), 
whereas for unbound magnons the distance is usually much larger (Fig.~\ref{fig:sketch_results}d).
Thereby, we identify the high-energy band $D$ as being the excitation of
repulsively bound two-magnon states\cite{HalatiBernier2023}. For fields below $B_c$ but still very large (e.g. $B = 30~$T),
we obtain an excitation band at even higher energy, as denoted by $T$ above the dashed line in Fig.~\ref{fig:theo_results}a.
This band involves three flipped spins bound next to each other, with the average distance between the first and third
spins being very small\cite{HalatiBernier2023} (i.e. $<3.5$ sites, see Fig.~\ref{fig:sketch_results}f). 
Thus, this band corresponds to the excitations of repulsively bound three-magnons. 

\section*{Identification of repulsively bound magnons}
The very good agreement between experiment and theory enables the unambiguous identification of experimentally observed repulsively bound magnon states (c.f. Fig.~\ref{fig:sketch_results}g and \ref{fig:sketch_results}h). $D_\pi$ and $D_{\pi/2}$ correspond to the excitations of repulsively bound two-magnon pairs, and $T_{\pi/2}$
to the repulsively bound three-magnon states.
The $D_{\pi/2}$ and $D_\pi$ modes are at the maximum and minimum of the excitation band
at the quasi-momenta $q=\pi/2$ and $q=\pi$, respectively (marked by green circles in Fig.~\ref{fig:theo_results}a and \ref{fig:theo_results}b).
However, the spectral weight corresponding to the repulsively bound three-magnon state is mostly concentrated around $q=\pi/2$ (Fig.~\ref{fig:theo_results}a).

The experimental observation of the repulsively bound magnon states highlights their important role in one-dimensional 
quantum many-body spin dynamics. 
These peculiar states appear as high-energy many-body dynamical features in a regime characterized by strong one-dimensional spin fluctuations, $0 \ll B<B_c$, where the antiferromagnetic exchange and the Zeeman interaction compete.
The contribution of the repulsively bound magnons to the dynamics is enhanced by the staggered effective field (Extended Data Fig.~\ref{fig:staggering_40T} and Extended Data Fig.~\ref{fig:staggering_30T}), a distinctive feature of BaCo\textsubscript{2}V\textsubscript{2}O\textsubscript{8} facilitating their experimental observation.
Above the critical field, quantum fluctuations are strongly suppressed and the spin dynamics is dominated by low-lying 
single-magnon excitations. There, the dynamical response of the magnon bound states is negligibly small and experimentally undetectable.

\begin{figure}
\centering
\includegraphics[width=0.8\textwidth]{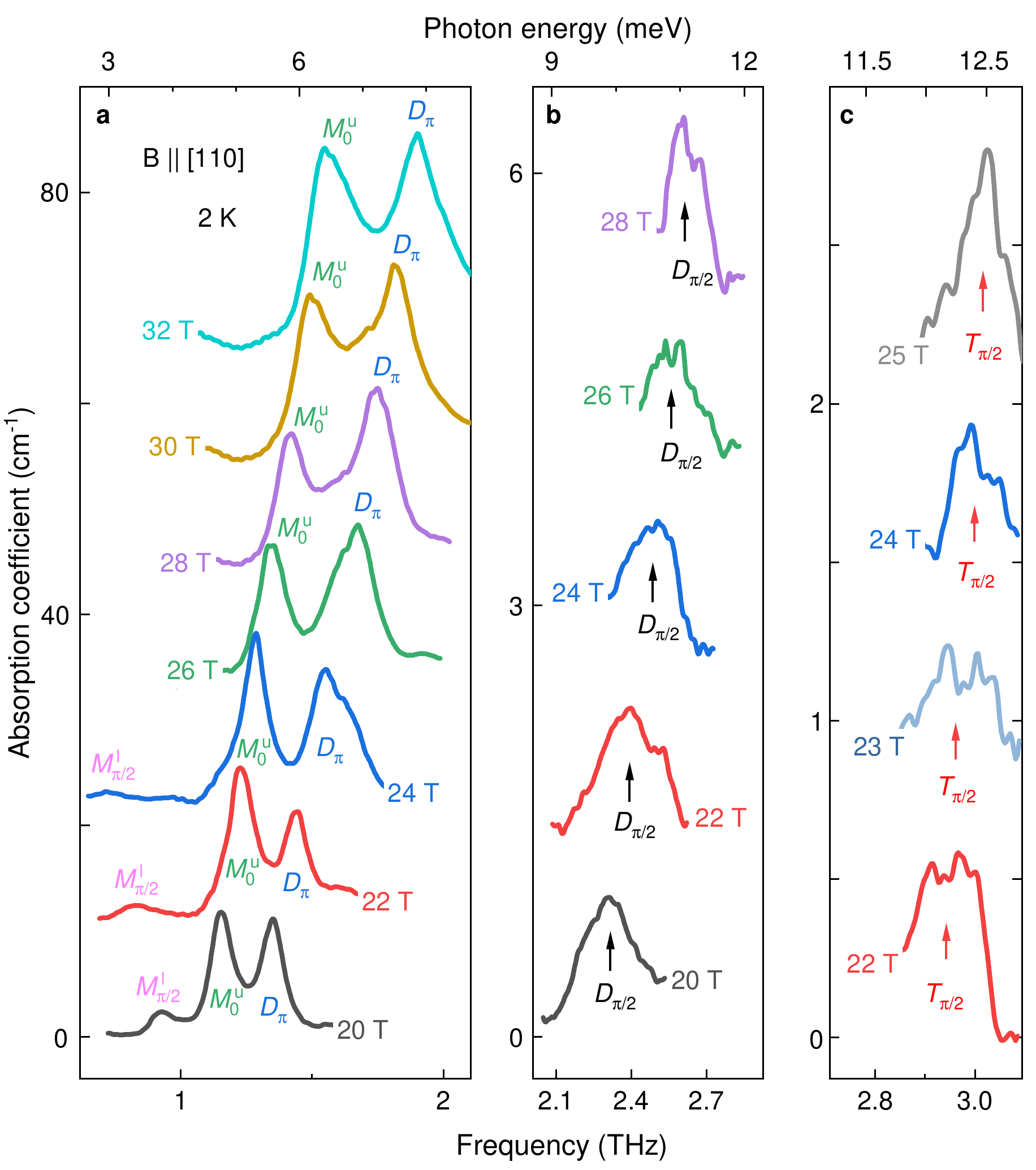}
\caption{\textbf{New types of high-energy excitations in the regime $0\ll B< B_c$}. \textbf{a}, Three different excitations
  $M^u_0$, $D_\pi$, and $M^l_{\pi/2}$ are observed, each with a different characteristic field dependence.
  \textbf{b}, Higher-energy modes $D_{\pi/2}$ and \textbf{c}, $T_{\pi/2}$ with smaller absorption coefficient, as indicated by the arrows,
  are resolved in high fields below $B_c$ (see also Extended Data Fig.~\ref{fig:rawdata}). While the $M^l_{\pi/2}$ mode softens, eigenenergies of the other
  modes increase with increasing fields. The spectra in higher fields are shifted upwards for clarity.}
\label{fig:exp_results_2}
\end{figure}

The characteristic lifetime of the repulsively bound states can be estimated from the experimentally resolved 
linewidth (Fig.~\ref{fig:exp_results_2}), which yields a typical value of $8\pm 2$ picoseconds. This is sufficiently long in comparison
with their relatively high characteristic frequencies ($\omega > 1.2~$THz, see Fig.~\ref{fig:exp_results_2}), allowing their experimental identification even in presence of dissipation. 
Such high frequencies correspond to the strong exchange interactions ensuring that the observed spin dynamics is governed by one-dimensional many-body quantum-mechanical effects rather than dissipation or thermal fluctuations.
At elevated temperatures, thermally enhanced fluctuations (e.g. lattice vibrations and spin fluctuations) will substantially reduce the lifetime of the repulsively bound magnons correspoding to broadened excitation lineshapes.

In conclusion, we demonstrate the existence of repulsively bound two- and three-magnon states in a solid-state compound, beyond the realm of ultracold atoms in optical lattices\cite{WinklerZoller2006, DeuchertCederbaum2012}, although thermal fluctuations and dissipative channels are present in addition to strong quantum fluctuations.
These unconventional repulsively bound states together with their transport properties are of particular interest in the fields of quantum communication, manipulation, and sensing\cite{Subrahmanyam2004,BarmanWinklhofer2021, YuanYan2022}.
Our results also motivate comprehensive studies of exotic high-energy magnon dynamics in quantum magnets and of 
general composite excitations in complex many-body systems.

\newpage

\begin{figure}
\centering
\includegraphics[width=0.95\textwidth]{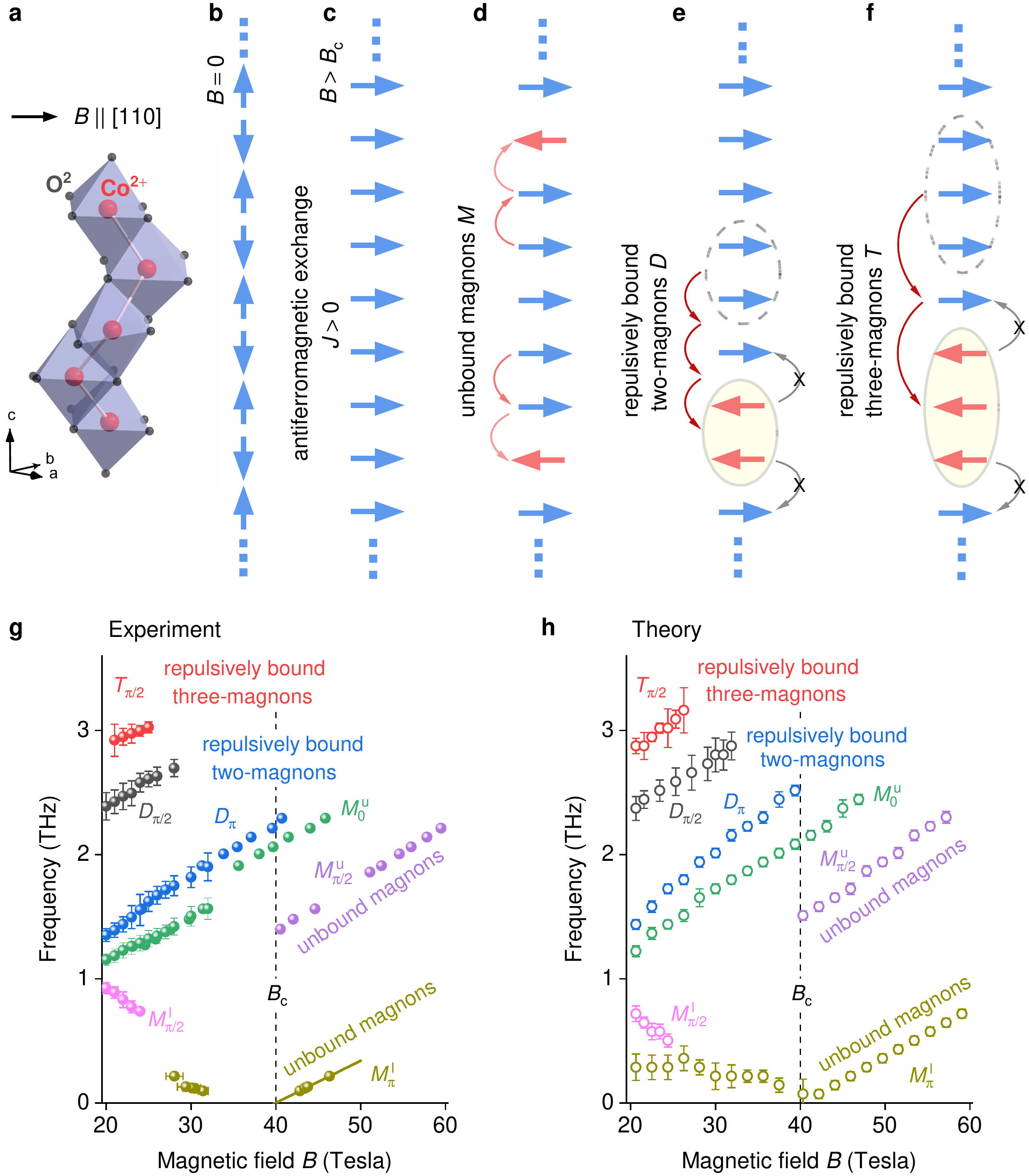}
\end{figure}

\begin{figure}[t!]
\caption{\textbf{Repulsively bound two- and three-magnons for a Heisenberg-Ising antiferromagnetic chain in strong transverse field}.
\textbf{a}, A four-fold screw chain of spin-$1/2$ Co\textsuperscript{2+} ions in BaCo\textsubscript{2}V\textsubscript{2}O\textsubscript{8}. 
\textbf{b}, Antiferromagnetic ground state in zero field corresponding to antiparallel alignment of spins with Ising easy axis in the
crystallographic \textit{c} direction. 
\textbf{c}, Fully field-polarized state above the 1D transverse-field Ising quantum critical point, $B>B_c$. 
\textbf{d}, States of two unbound magnons corresponding to two spin flips (red arrows) which are well separated and propagate independently in the chain (red curved arrows). 
\textbf{e}, Sketch of repulsively bound two-magnon states with the two spin flips occupying adjacent sites.  
\textbf{f}, Sketch of repulsively bound three-magnon states with the three spin flips occupying adjacent sites.
The bound magnons do not propagate separately (grey curved arrows) but as an entity (red curved arrows) along the chain.
  \textbf{g}, Eigenfrequencies as a function of transverse field for all the experimentally observed magnetic excitations
  and \textbf{h}, for the corresponding modes obtained by the numerically exact tMPS algorithm using the parameters
  $J=2.82~$meV, $\Delta=1.92$, $g^x_u = 3.06$, $g^x_s = 0.66$, and $g^z = 0.21$.
  Very good agreement between experiment and theory is achieved. Above $B_c=40~$T the single-magnon
  modes ($M_\pi^l$, $M_{\pi/2}^u$ and $M_0^u$ at the quasi-momenta $q=\pi$, $\pi/2$ and $0$, see Fig.~\ref{fig:theo_results}) are observed in the field-polarized phase. 
  The solid line in (\textbf{g}) is a linear fit of the low-lying $M_\pi^l$ mode which determines the critical field
  of $B_c=40~$T. Below $B_c$, excitations of the repulsively bound three magnons ($T_{\pi/2}$ at $q=\pi/2$),
  two magnons ($D_{\pi/2}$ and $D_\pi$ at $q=\pi/2$ and $q=\pi$), as well as the single magnons
  ($M_{\pi/2}^l$ and $M_\pi^l$ at $q=\pi/2$ and $q=\pi$) are identified by their field dependent
  eigenenergies. Experimental and theoretical linewidths are indicated by bars.}
  \label{fig:sketch_results}
\end{figure}

\begin{figure}[t!]
\centering
\includegraphics[width=0.6\textwidth]{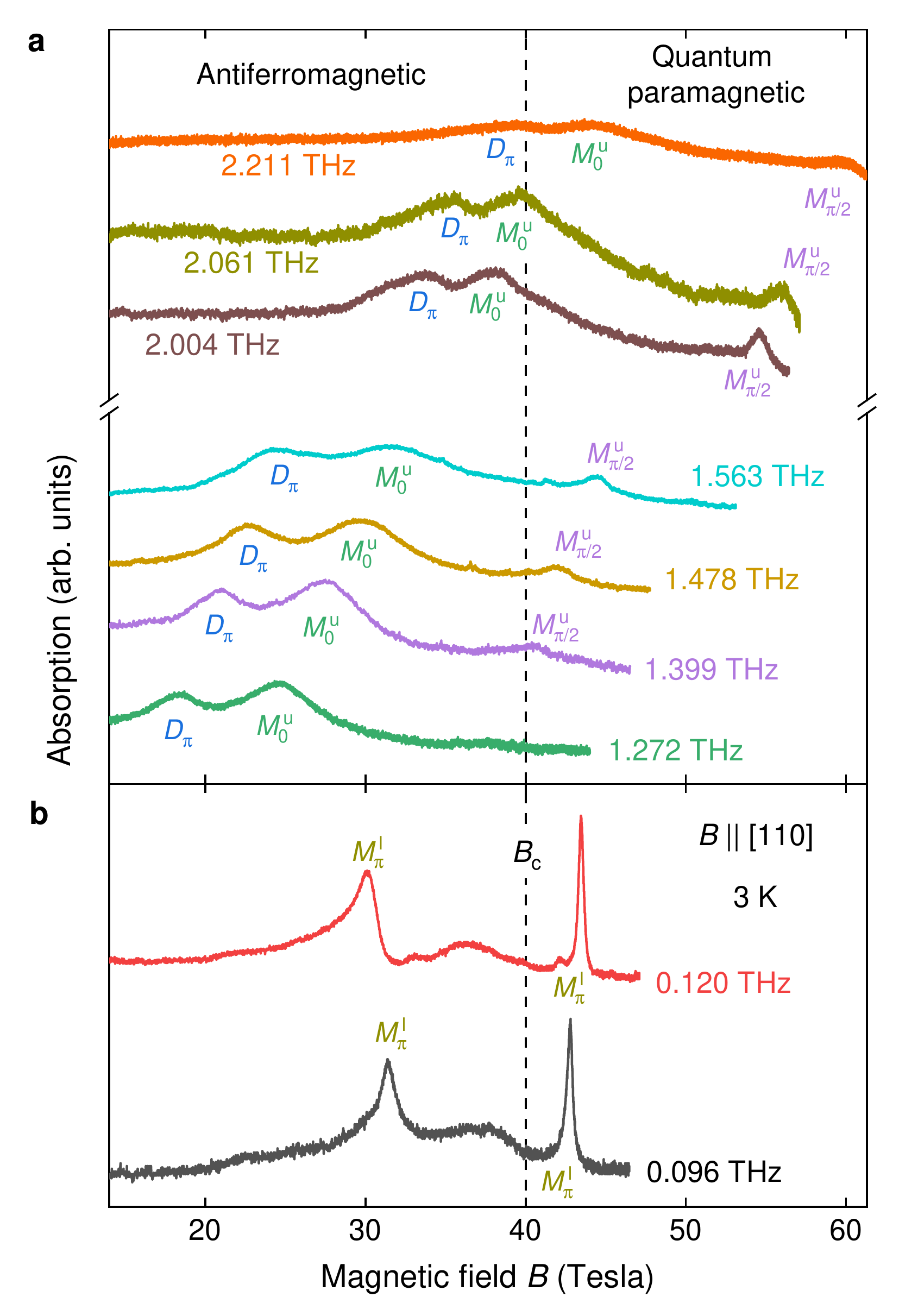}
\caption{\textbf{Quantum spin dynamics around the one-dimensional transverse-field Ising quantum critical point}. \textbf{a}, Absorption spectra at various frequencies exhibit resonance peaks corresponding to the higher-energy excitations $D_\pi$, $M^u_0$, and $M^u_{\pi/2}$.
  \textbf{b}, Field-dependent evolution of the lowest-lying mode $M_\pi^l$ indicates the gap-closing and reopening in the vicinity of
  the quantum phase transition at $B_c = 40$~T, which belongs to the one-dimensional transverse-field Ising universality
  class\cite{WangLoidl2018b}.  Above $B_c$ the system enters a fully field-polarized quantum paramagnetic phase (see Fig.~\ref{fig:sketch_results}c for an  illustration). The spectra of higher frequencies are shifted upwards for clarity.}
\label{fig:exp_results_1}
\end{figure}

\clearpage

\begin{figure}
\centering
\includegraphics[width=1\textwidth]{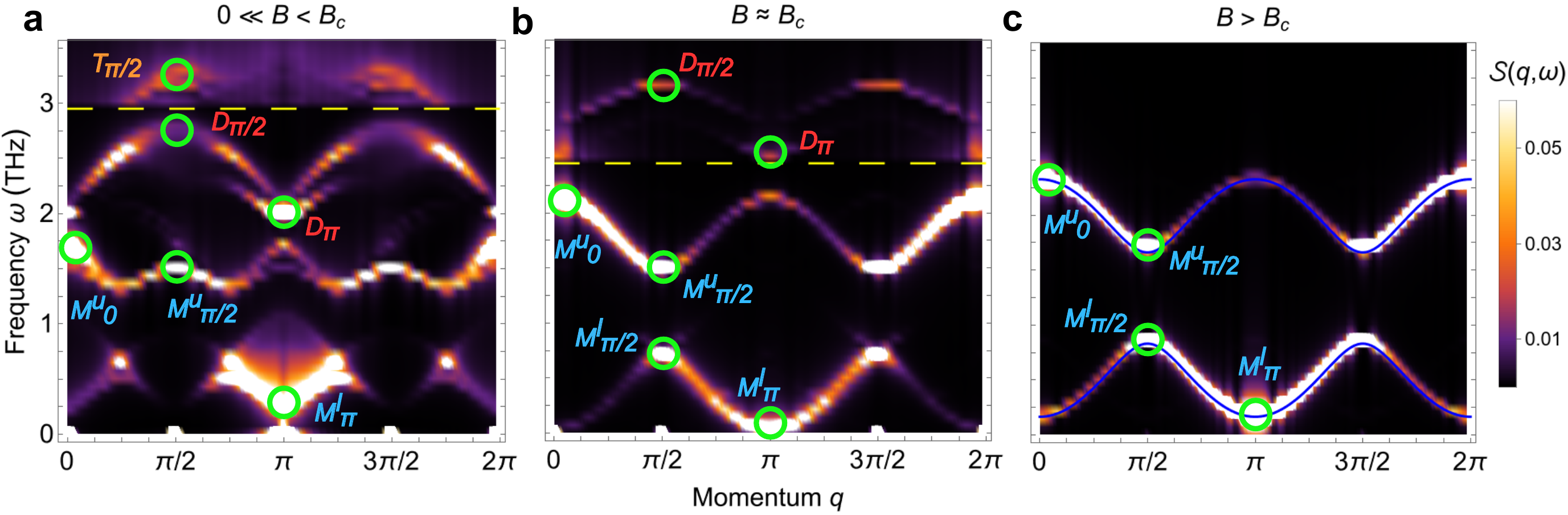}
\caption{\textbf{Characteristic quantum spin dynamics}.
Dynamical spin structure factor $\mathcal{S}(q,\omega)$ as a function of momentum $q$ and frequency $\omega$ for an applied magnetic field of \textbf{a}, $0 \ll B = 30$~T~$< B_c$, \textbf{b}, $B = 40.3$~T~$\approx B_c$, and \textbf{c}, $B = 45$~T~$>B_c$,
  using the numerically exact time-dependent matrix-product-state algorithm\cite{Daley2014, WhiteFeiguin2004, Schollwoeck2011}
  for the transverse-field Heisenberg-Ising model in Eq.~(\ref{eq:Hamiltonian}). 
  The green circles mark the modes at the quasi-momenta $q=0$, $\pi/2$, and $\pi$ in the extended Brillouin zone, which correspond to zero-momentum transfer in the reduced zone scheme due to the four-fold screw symmetry in the spin-chain structure (see Fig.~\ref{fig:sketch_results}a) and is experimentally detectable by terahertz spectroscopy.
    \textit{M}, \textit{D}, and \textit{T} denote the excitations of unbound magnons, repulsively bound two-magnons, and
  repulsively bound three-magnons, respectively. In the corresponding labels the subscript denotes the momentum $q$.
  In each panel we normalized $\mathcal{S}(q,\omega)$ to its maximum value. To highlight the high-frequency modes
  above the horizontal dashed lines, we multiplied $\mathcal{S}(q,\omega)$ with a factor of 10. The blue curves in (\textbf{c})   depict the one-magnon dispersion computed analytically\cite{HalatiBernier2023}.
}
\label{fig:theo_results}
\end{figure}

\clearpage

\newpage

\begin{methods}

\subsection{Sample preparation and characterization.}

High quality single crystals of BaCo\textsubscript{2}V\textsubscript{2}O\textsubscript{8} were
synthesized by a solid-state reaction using a mixture of BaCO\textsubscript{3}
(99+\% Merck), Co\textsubscript{3}O\textsubscript{4} (99.5\% Alfa Aesar) and V\textsubscript{2}O\textsubscript{5}
(99.5\% Strem Chemicals) in a $1:\frac{2}{3} :1$ molar ratio. 
A four-mirror image furnace (FZ-T-10000-H-VI-VP, Crystal Systems Inc.) was used for the
single-crystal growth by means of the floating-zone technique\cite{NiesenLorenz2013}. 
For the optical experiments, single crystals were oriented at room
temperature using x-ray Laue diffraction. Samples were cut perpendicular to the crystallographic [110]
direction with a typical size of 4 $\times$ 4 $\times$ 1 mm\textsuperscript{3}.

\subsection{Magnetization measurements in pulsed magnetic fields.}

A bar-shaped BaCo\textsubscript{2}V\textsubscript{2}O\textsubscript{8} single crystal of 2 mm length along
the [110] direction and a cross section of about 1 mm\textsuperscript{2} was used for the magnetization
measurements in a pulsed magnetic field which was applied along the [110] direction, i.e. $B \parallel [110]$. The
pulsed field had a rise time of 7 ms and a pulse duration of 20 ms.

The magnetization of BaCo\textsubscript{2}V\textsubscript{2}O\textsubscript{8} obtained at 1.8 K is presented in Extended
Data Fig.~\ref{fig:magnetization} as a function of the applied transverse magnetic field, $B \parallel [110]$. 
The magnetization increases continuously with magnetic field, and exhibits a saturation feature at the quantum phase
transition of $B_c = 40~\text{T}$, as confirmed
by the measurements of sound velocity and magnetocaloric effect\cite{WangLoidl2018b}. Above $B_c$ a slight increase
of the magnetization is ascribed to a van Vleck paramagnetic contribution with a susceptibility $ 0.00895 \mu_B/\text{T}$ (dotted line). 
As shown by the solid line, the experimental data are very well described by the sum of the van Vleck contribution
(dotted line) and the density matrix renormalization group results for the one-dimensional Heisenberg-Ising model (dashed line) at
zero temperature, with the same parameters as for the observed magnetic excitations (see Fig.~\ref{fig:sketch_results} and Fig.~\ref{fig:theo_results}).

\subsection{Terahertz spectroscopy in pulsed and static magnetic fields.}

The high-field electron-spin-resonance spectroscopy in a pulsed magnetic field up to 61~T were performed
at Helmholtz-Zentrum Dresden-Rossendorf (HZDR). Absorption spectra of quasi-monochromatic electromagnetic waves
were measured as a function of the pulsed magnetic field for frequencies above 1.2~THz using a free electron
laser\cite{ZvyaginSeidel2009} and below 0.6~THz using VDI microwave sources (Virginia Diodes Inc, USA) at HZDR.
The terahertz broad-band transmission measurements were performed in static magnetic fields up
to 32~T using a Bitter electromagnet at the High Field Magnet Laboratory in Nijmegen. Terahertz
electromagnetic waves were generated by a Mercury lamp and detected by a silicon bolometer. 
The spectra were recorded using a Fourier-transform spectrometer Bruker IFS-113v. For all these measurements,
the magnetic fields were applied along the crystallographic [110] direction.

\subsection{Anisotropic $g$-factors and effective magnetic field in BaCo\textsubscript{2}V\textsubscript{2}O\textsubscript{8}.}

Since the CoO\textsubscript{6} octahedra in BaCo\textsubscript{2}V\textsubscript{2}O\textsubscript{8} are slightly
distorted with the apical Co-O bonds tilted off the $c$-axis (i.e. the Ising-like axis, see Fig.~\ref{fig:sketch_results}a), the magnetic local principal axes do not coincide with the crystallographic axes. This results in staggered Landé
$g$-factors\cite{KimuraWatanabe2013, NiesenLorenz2013}
\begin{align}
\label{eq:bacovo_0direction}
g^{xx}_j &= \qty(g_1\cos^2\theta+g_2\sin^2\theta)\cos^2\qty(\frac{\pi}{2}(j-1)) + g_3\sin^2\qty(\frac{\pi}{2}(j-1)), \\
g^{xy}_j &= 0, \nonumber\\
g^{xz}_j &= (g_2 - g_1)\cos\theta\sin\theta\cos\qty(\frac{\pi}{2}(j-1)), \nonumber \\
g^{yy}_j &= \qty(g_1\cos^2\theta+g_2\sin^2\theta)\sin^2\qty(\frac{\pi}{2}(j-1)) + g_3\cos^2\qty(\frac{\pi}{2}(j-1)), \nonumber \\
g^{zz}_j &= g_1\sin^2\theta+g_2\cos^2\theta, \nonumber \\
g^{yz}_j &= (g_2 - g_1)\cos\theta\sin\theta\sin\qty(\frac{\pi}{2}(j-1)), \nonumber
\end{align}
where $\theta=5^\circ$ is the tilt angle from the $c$-axis and $g_1$, $g_2$, and $g_3$ are the values of
the $g$-tensor along the magnetic principle axes. The $x$-coordinate is defined along the crystallographic
[110] direction, parallel to the applied magnetic field, while the $y$- and $z$-coordinates are along the
[-110] and [001] directions, respectively.

For a transverse magnetic field in the $x  \parallel \text{[110]}$ direction, the Zeeman term due to the effective
magnetic field for a single chain in BaCo\textsubscript{2}V\textsubscript{2}O\textsubscript{8} is given
by $\mu_B B_x \sum_j\left( g^{xx}_j S_j^x+g^{xz}_j S_j^z \right)$, where $g^{xx}_{1,2,3,4}=(3.72,~2.4,~3.72,~2.4)$
and $g^{xz}_{1,2,3,4}=(0.21,~0,~-0.21,~0)$, corresponding to $g_{1,2,3}=3.7,~6.1,~\text{and}~2.4$.
Due to the site-dependent anisotropic $g$-factors, the effective magnetic field in the $x$-direction has
a uniform and a staggered component, corresponding to $g^{xx}_j=g^x_u-(-1)^j g^x_s$, with $g^x_u=3.06$ and $g^x_s=0.66$. 
In the $z$-direction the effective field has a four-fold periodicity, $g^{xz}_j=g^z \sin(\frac{\pi}{2}j)$, with a small value of $g^z=0.21$, whose contribution does not play a crucial role for the existence of the repulsively bound states\cite{HalatiBernier2023}. 
The parameters obtained are summarized in Extended Data Table~1 and Extended Data Table~2, and compared with the reported values in the literature. While the values of $J$, $\Delta$, and $g^x_u$ are consistent with the reported values\cite{KimuraWatanabe2013,FaureGrenier2018,WangLoidl2019,OkutaniHagiwara2021,wang2023spin}, the value of $g^x_s$ corresponding to the staggered field is relatively small and was neglected in the literature. In contrast, we have shown the importance of a finite staggered field, which is responsible for the splitting of the low-lying single-magnon band (see Extended Data Fig.~\ref{fig:staggering_45T}) and enhances the dynamical response of the repulsively bound magnons (see Extended Data Fig.~\ref{fig:staggering_40T} and Extended Data Fig.~\ref{fig:staggering_30T}). These parameters are further constrained by the field dependence of the magnetization (see Extended Data Fig.~\ref{fig:magnetization}).

\subsection{Numerical simulations using the matrix product state (MPS) technique.}

We investigate theoretically the excitations observed experimentally by computing the dynamical spin structure
factor, $\mathcal{S}(q,\omega)$. Within linear response, the definition we employ is given by 
\begin{align}
\label{eq:structure_factor}
\mathcal{S}(q,\omega) =\sum_{l=L/2-1}^{L/2+2}&\Bigg[ \left|(gS)^{zz}_l(q,\omega)\right|^2+\left|(gS)^{yy}_l(q,\omega)\right|^2
+\left|(gS)^{zy}_l(q,\omega)\right|^2+\left|(gS)^{yz}_l(q,\omega)\right|^2\Bigg].
\end{align}
Since the terahertz field is unpolarized and perpendicular to the applied external field, Eq.~(\ref{eq:structure_factor})
contains the relevant transverse components which are defined by
\begin{align}
  (gS)^{\alpha \beta}_l(q,\omega) =\frac{1}{\sqrt{L}}\int_0^\infty \sum_{j=1}^L e^{i(\omega t-qj)}\mathcal{G}^\alpha_j \mathcal{G}^\beta_lS^{\alpha \beta}_{j,l}(t) dt
\end{align}  
with $\alpha,\beta\in\{x,y,z\}$, $\mathcal{G}^x_l=g^{zx}_l$, $\mathcal{G}^y_l=g^{yy}_l+g^{zy}_l$, $\mathcal{G}^z_l=g^{zz}_l+g^{yz}_l$, 
and the two-point correlation functions $S^{\alpha \beta}_{j,l}(t)=\bra{0}S^{\alpha }_{j}(t)S^{\beta}_{l}\ket{0}$ for the spin operators $S^\alpha_j$
and $S^\beta_j$. The index $l$ enumerates the four Co$^{2+}$ spin sites within the central unit cell. 
The ground state, denoted by $\ket{0}$, and the time-dependent correlations are computed numerically
using MPS algorithms (see below).

The numerical simulations performed in this work for the one-dimensional model, Eq.~(\ref{eq:Hamiltonian}),
are based on the matrix product states techniques\cite{Daley2014, WhiteFeiguin2004, Schollwoeck2011}. 
This method is variational in the space of matrix product states characterized by a certain matrix dimension, called the bond dimension.
The loss of accuracy in the representation of the wave function within an approximation state is quantified by the truncation error.
The ground state $\ket{0}$ is obtained using a finite-size density matrix renormalization group algorithm\cite{Schollwoeck2011}
in the MPS representation implemented using the ITensor Library\cite{FishmanStoudenmire2020}. The convergence is
ensured by a maximal bond dimension up to 300, for which the truncation error is at most $10^{-12}$. 
The two-point correlations are computed numerically using the time-dependent MPS method\cite{Daley2014, WhiteFeiguin2004, Schollwoeck2011}.
We considered systems of $L = 124$ sites and bond dimensions up to 300. This ensures that at the final
time $tJ/\hbar = 110$ the truncation error is $\leq 10^{-7}$ (or $\leq 10^{-10}$ above the phase transition).
The time step is $\delta t J/\hbar=0.05$, and the measurements were performed every fourth time step. In order
to minimize the numerical artifacts arising due to the use of open boundary conditions, we applied a Gaussian filter,
$f(j)=e^{-4\left(1-\frac{2j}{L-1}\right)^2}$, to the dynamic correlations before performing the numerical Fourier transform, here
$j$ labels the sites.

\subsection{Effects of the effective staggered magnetic field.}

We have emphasized that the staggered magnetic field plays a very important role in the visibility and detection of
the multi-magnon bound states.
This can be systematically shown by performing numerically
exact calculations with the MPS technique for various values of $g^x_s$, which is responsible for the staggered field.

We consider three different regimes: above the phase
transition, $B>B_c$ (see Extended Data Fig.~\ref{fig:staggering_45T}), at the phase transition $B\approx B_c$
(see Extended Data Fig.~\ref{fig:staggering_40T}), and below (but still at large magnetic field), $0\ll B<B_c$
(see Extended Data Fig.~\ref{fig:staggering_30T}). The values of the other parameters are the same as for
the results of the main text, $J = 2.82~\text{meV}$, $\Delta = 1.92$, $g^x_u= 3.06$, and $L = 124$, except the four-fold
field which is absent $g^z=0$. We can show that the small value
of $g^z=0.21$ does not influence the characteristic features of the respulsively bound magnon states (see Ref.~\cite{HalatiBernier2023}).

For $B>B_c$, the spins are polarized along the field direction. Without a staggered field, $g^x_s=0$, the excitations
are characterized by the dynamics of a single magnons moving through the chain. This determines a single
cosine-shaped band in the structure factor (see Extended Data Fig.~\ref{fig:staggering_45T}a, denoted by \textit{M})
If the staggering becomes finite, $g^x_s>0$, this band is split into two with a gap. 
This is because that a propagating single magnon around the lattice experiences site-dependent lower or higher effective magnetic field.
This can also be seen from the fact that the gap increases with increasing staggered field
(see Extended Data Fig.~\ref{fig:staggering_45T} b-d for $g^x_s\in\{0.31,0.66,0.94\}$). 
We mention that the results in Extended Data Fig.~\ref{fig:staggering_45T}c corresponds to the same
staggering, $g^x_s=0.66$, as for the parameters considered in Fig.~\ref{fig:theo_results}
describing BaCo\textsubscript{2}V\textsubscript{2}O\textsubscript{8}.

In the case of a magnetic field close to the critical value, $B\approx B_c$, we can clearly
distinguish the two-magnon bound states features at the experimental value $g^x_s=0.66$ (Extended Data Fig.~\ref{fig:staggering_40T}c,
see also Fig.~\ref{fig:theo_results}). If we increase the staggering, the spectral weight of the repulsively bound
two-magnon excitations becomes even more prominent (Extended Data Fig.~\ref{fig:staggering_40T}d).
Whereas at lower and vanishing staggering, the spectral weight strongly decreases and becomes hardly
discernible (see Extended Data Fig.~\ref{fig:staggering_40T}b and \ref{fig:staggering_40T}a).

Similarly, for magnetic field below $B_c$ but still large, i.e. $0 \ll B<B_c$ , starting from the experimental
value $g^x_s= 0.66$ (Extended Data Fig.~\ref{fig:staggering_30T}c), an increase or decrease of the staggered
field will enhance or reduce, respectively, the spectral weight of
the repulsively bound two-magnon (denoted by \textit{D}) and three-magnon (denoted by \textit{T}) states. 
It is important to note that even without the staggered field the repulsively bound multi-magnon states
are present in the model, due to the interplay of the antiferromagnetic interactions and strong magnetic
fields\cite{HalatiBernier2023}. However, their presence in the dynamical structure 
factors as well-defined features with a spectral weight large enough to be detected experimentally is due to the presence of a strong staggering of the effective magnetic field, as it is the case of BaCo\textsubscript{2}V\textsubscript{2}O\textsubscript{8}.

\subsection{Guiding principles for identifying repulsively bound magnons.}

Our study reveal the following guiding principles for identifying repulsively bound magnons. 
First, spin interactions should be dominated by antiferromagnetic intrachain exchange with negligible interchain 
couplings, enhancing one-dimensional quantum fluctuations.
Second, the intrachain interaction should be sufficiently stronger than thermal energy, such that the spin dynamics 
is governed by quantum mechanical effects. A strong intrachain interaction also corresponds to a large 
quantum critical field, therefore the three-dimensional order stabilized by the interchain couplings is already 
suppressed well below the one-dimensional quantum critical point. 
Third, to avoid a fast decay of the repulsively bound magnons, dissipation channels (e.g. a direct 
coupling with phonons or other degrees of freedom) should be minimized as much as possible.
Last but not least, the staggered magnetic field in BaCo\textsubscript{2}V\textsubscript{2}O\textsubscript{8} 
increases the separation of the repulsively bound magnons from the lower-energy excitations, and also enhances 
their dynamical responses, facilitating the experimental identification.

\end{methods}

\clearpage

\bibliographystyle{naturemag}


\newpage

\begin{addendum}
\item We thank M. Garst, T. Giamarchi, S. Wolff, J. Wu, and H. Zou for helpful discussions.
We acknowledge support by the European Research Council (ERC) under the Horizon 2020 research and innovation programme, grant
agreement No. 950560 (DynaQuanta), by the Natural Sciences and Engineering Research Council of Canada (NSERC)
[funding references No. RGPIN-2021-04338 and No. DGECR-2021-00359], and by the Swiss National Science Foundation under Division II grants 200020-188687 and 200020-219400. 
This research was supported in part by the National Science Foundation under Grants No. NSF PHY-1748958 and PHY-2309135.
We acknowledge funding from the Deutsche Forschungsgemeinschaft (DFG, German Research Foundation)
under project number 107745057 - TRR 180 (F5), project number 277146847 - CRC 1238 (A02,B01,B05,C05), project number 277625399 - TRR 185 (B4), project number 247310070 - SFB 1143, Project number 511713970 - CRC 1639, and project number 390534769 - EXC 2004/1 Cluster of Excellence Matter and Light for Quantum Computing (ML4Q).
We also acknowledge the support of the HFML-RU/FOM and the HLD at Helmholtz-Zentrum Dresden-Rossendorf (HZDR), members of the European Magnetic Field Laboratory (EMFL). Parts of this
research were carried out at ELBE at the HZDR, a member of the Helmholtz Association. 

\item[Author contributions]
Z.W. conceived the experiment and coordinated the project.
C.M.H., J.S.B. and C.K. performed the theoretical analysis.
Z.W., A.P., J.M.K. and S.Z. carried out the spectroscopic measurements.
D.I.G., T.L. and Z.W. did the magnetization measurements.
S.N., O.B. and T.L. prepared and characterized the single crystals.
Z.W., C.M.H., J.S.B. and C.K. analysed the data and interpreted the results.
Z.W., C.M.H., J.S.B. and C.K. wrote the manuscript with input from T.L. and A.L. 
All authors commented on the manuscript.

\item[Competing interests] The authors declare no competing interests.

\item[Data availability]
All data needed to evaluate the conclusions in the paper are included in this paper. Additional data that support the plots and other analysis in this work are available from the corresponding author upon request.

\end{addendum}

\newpage

\renewcommand{\figurename}{Extended Data Fig.}
\setcounter{figure}{0}

\renewcommand{\tablename}{Extended Data Table}

\begin{figure}
\centering
\includegraphics[width=0.40\textwidth]{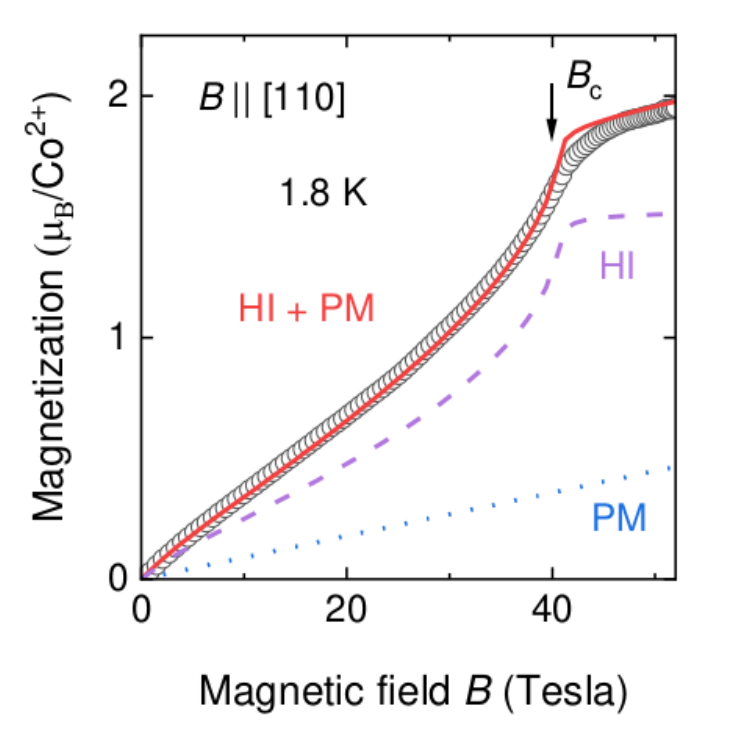} 
\caption{Magnetization of BaCo\textsubscript{2}V\textsubscript{2}O\textsubscript{8} as a function
  of the applied magnetic field along the crystallographic [110] direction\cite{WangLoidl2018b}, i.e. $B \parallel [110]$.
  The solid line shows the theoretical result, as a sum of a paramagnetic (PM) van Vleck contribution
  (dotted line) and the contribution of the one-dimensional Heisenberg-Ising (HI) model in Eq.~(\ref{eq:Hamiltonian}) (dashed line).}
\label{fig:magnetization}
\end{figure}

\begin{table}
\begin{center}
\caption{The parameters of $J$ and $\Delta$.}
\vspace{4pt}
\setlength\tabcolsep{14pt}

\begin{tabular}{lllllll}
\hline
\hline
   		&Ref.\cite{KimuraWatanabe2013}   &Ref.\cite{FaureGrenier2018}  &Ref.\cite{WangLoidl2019} &Ref. \cite{OkutaniHagiwara2021} &Ref.\cite{wang2023spin} &This work \\
 \hline
 $J$ (meV)  &2.58  &3.07 &2.6 &2.58  &2.67 &2.82 \\
 \hline
 $\Delta$   &2.17  &1.89 &2.17 &2.17  &2.17 &1.92 \\
 \hline
 \hline
\end{tabular}
\end{center}
\label{table1}
\end{table}


\begin{table}
\sffamily
\begin{center}
\caption{The parameters of the \textit{g}-values $g_u^x$, $g_s^x$, and $g_z$.}
\vspace{4pt}
\setlength\tabcolsep{14pt}
\begin{tabular}{lllllll}
\hline
\hline
   		&Ref.\cite{KimuraWatanabe2013}   &Ref. \cite{OkutaniHagiwara2021}  &This work \\
  \hline
  \vspace{2pt}
 $g_u^x$   &2.95  &2.95 &3.06 \\
 \hline
 $g_s^x$  &0 &0 &0.66\\
 \hline
 $g_z$ &0.41 &0.41 &0.21\\
 \hline
 \hline
\end{tabular}
\end{center}
\label{table2}
\end{table}

\begin{figure}
\centering
\includegraphics[width=1\textwidth]{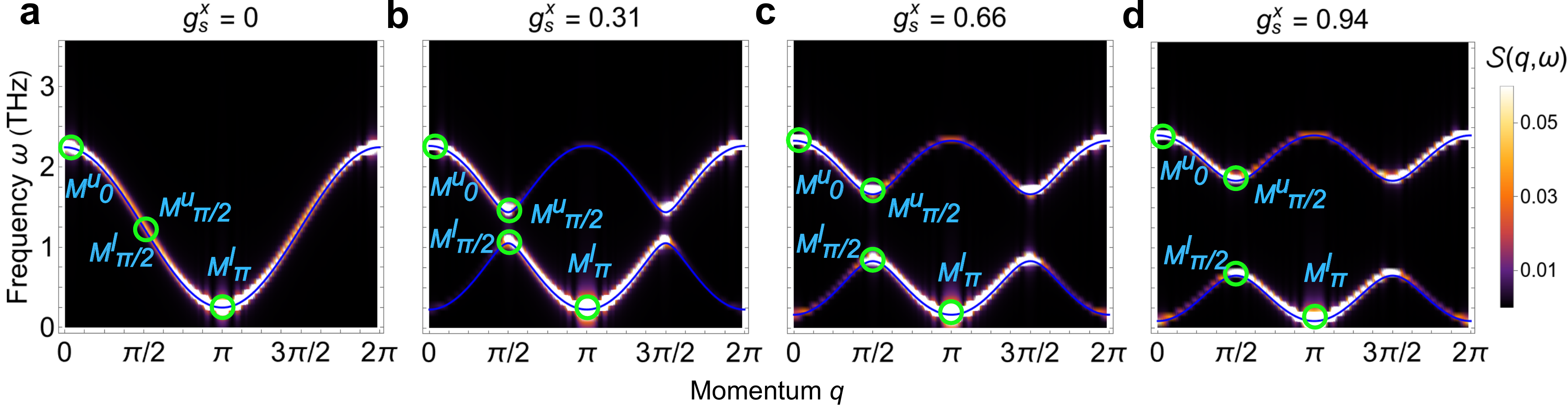} 
\caption{Dynamical spin structure factor $\mathcal{S}(q,\omega)$ as a function of momentum $q$ and
  frequency $\omega$ [see Eq.~(\ref{eq:structure_factor})] at an applied field of $B = 45~\text{T} > B_c$ for
  various values of $g^x_s$ corresponding to different effective staggering of the magnetic field. 
  \textbf{a}, Without a staggered field
  (i.e. $g^x_s=0$), the spin dynamics is characterized by a single cosine-shaped band of unbound-magnon excitations, labelled as \textit{M}. 
  \textbf{b}, \textbf{c}, and \textbf{d},
  With a finite staggered field (i.e. $g^x_s =$~0.31, 0.66, and 0.94, respectively) this band is split into two bands separated by a gap.
  The gap increases with increasing staggered field.  The data in (\textbf{c}) correspond to the experimental value of the
  staggering in BaCo\textsubscript{2}V\textsubscript{2}O\textsubscript{8} (see Fig.~\ref{fig:theo_results}c). The blue lines are analytical results for the
  one-magnon excitations\cite{HalatiBernier2023}.}
\label{fig:staggering_45T}
\end{figure}

\clearpage

\begin{figure}
\centering
\includegraphics[width=1\textwidth]{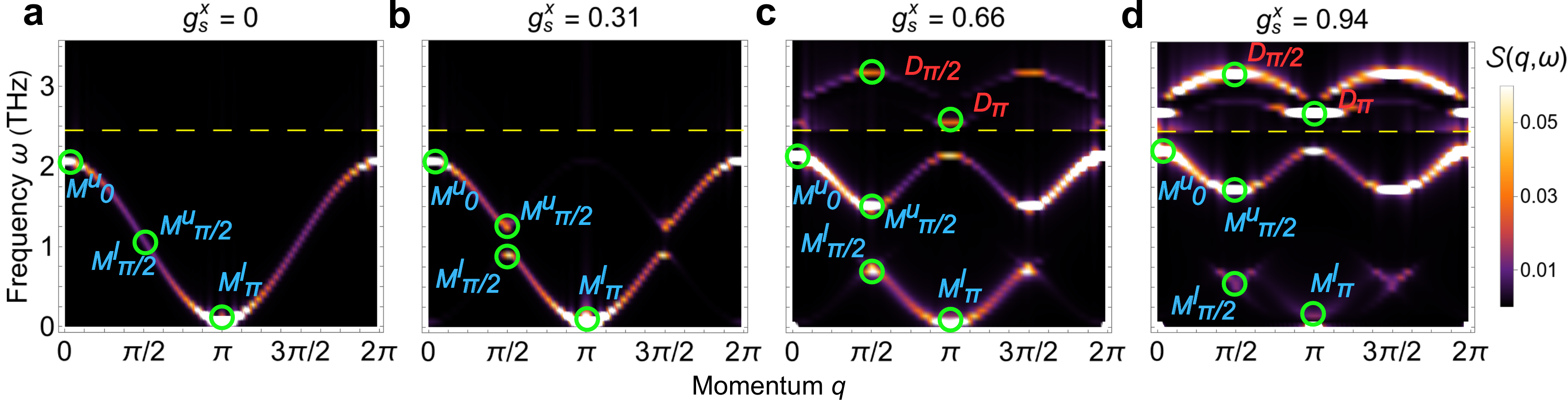} 
\caption{Dynamical spin structure factor $\mathcal{S}(q,\omega)$ as a function of momentum $q$ and
  frequency $\omega$ [see Eq.~(\ref{eq:structure_factor})] at an applied field
  of $B = 40.3~\text{T} \approx B_c$ for various values of $g^x_s$ corresponding to different
  effective staggering of the magnetic field.  Above the dashed line, the spectral weight of the repulsively
  bound two-magnon states (labelled as \textit{D}) is multiplied by a factor of 10. The data in (\textbf{c}) correspond to the experimental
  value of the staggering for BaCo\textsubscript{2}V\textsubscript{2}O\textsubscript{8} (see Fig.~\ref{fig:theo_results}b).}
\label{fig:staggering_40T}
\end{figure}

\begin{figure}
\centering
\includegraphics[width=1\textwidth]{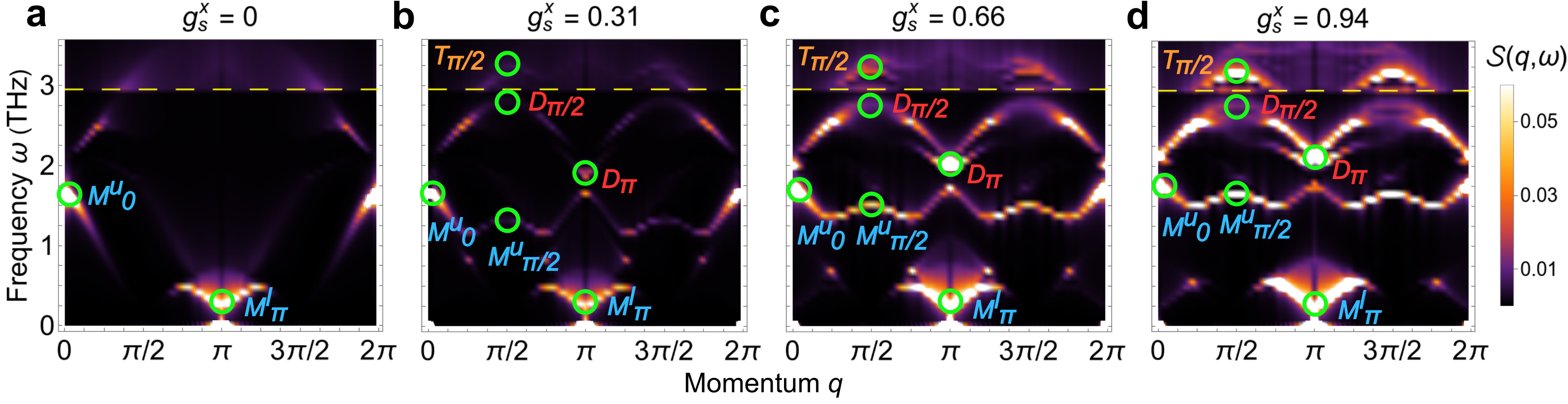} 
\caption{Dynamical spin structure factor $\mathcal{S}(q,\omega)$ as a function of momentum $q$ and
  frequency $\omega$ (see Eq.~(\ref{eq:structure_factor})) at an applied field of $B = 30~\text{T}$
  (i.e $0\ll B<B_c$) for various values of $g^x_s$ corresponding to different effective staggering
  of the magnetic field. 
  \textbf{a}, Without a staggered field (i.e. $g^x_s=0$), the high-energy features above the unbound-magnon
  excitation band are hard to identify. 
  However, with increasing staggered field \textbf{b}, for $g^x_s =$~0.31, \textbf{c}, for $g^x_s =$~0.66, and \textbf{d}, for $g^x_s =$~0.94, the spectral weight of the features
  of repulsively bound two-magnon and three-magnon excitations (labelled as \textit{D} and \textit{T}, respectively) can be clearly identified and is continuously enhanced. Above the dashed line, the spectral weight of the repulsively bound three-magnon states is multiplied by a factor of 10.
  The data in (\textbf{c}) correspond to the experimental value of the staggering
  for BaCo\textsubscript{2}V\textsubscript{2}O\textsubscript{8} (see Fig.~\ref{fig:theo_results}a).}
\label{fig:staggering_30T}
\end{figure}

\begin{figure}
\centering
\includegraphics[width=1\textwidth]{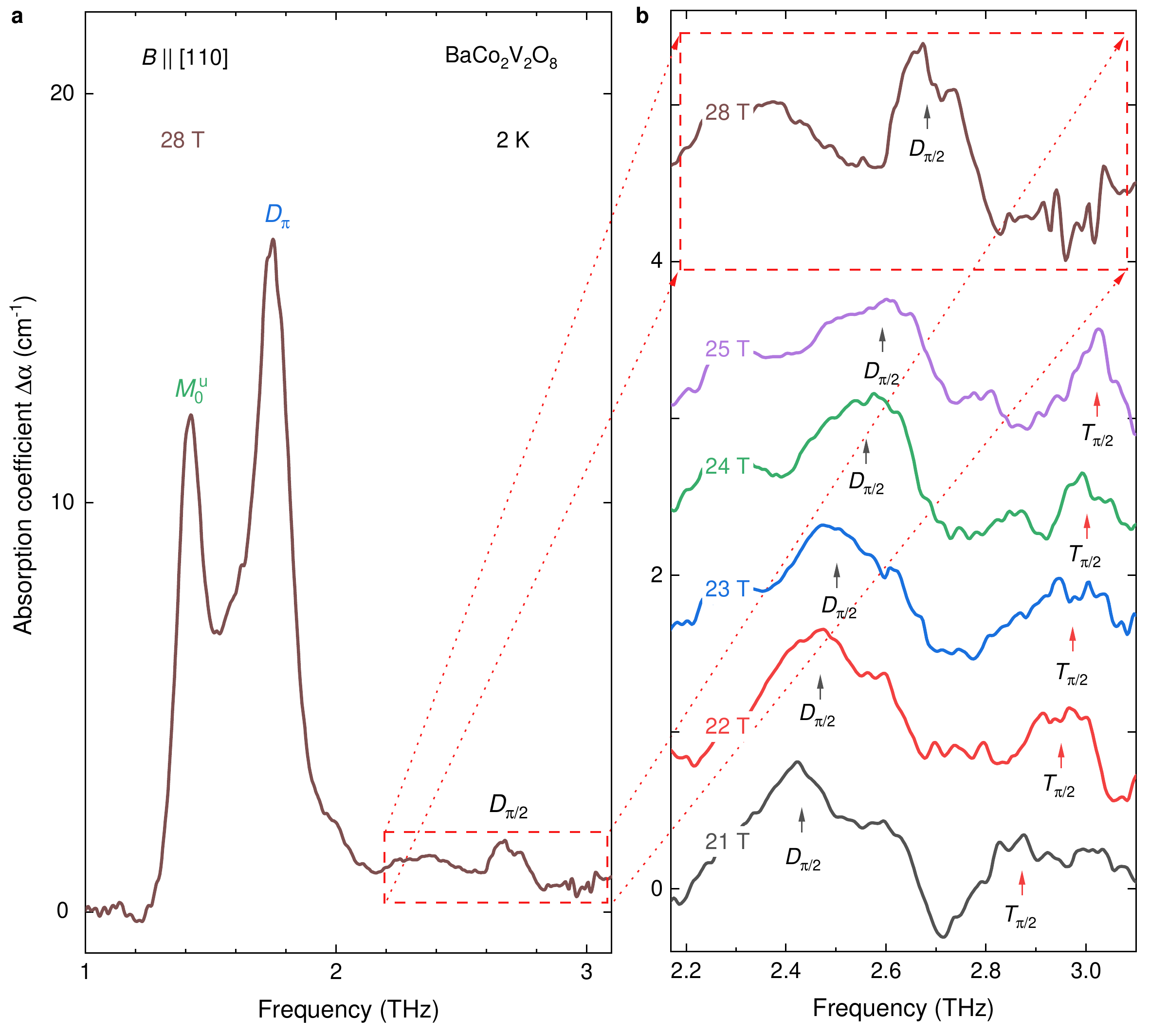} 
\caption{Absorption coefficient measured in static magnetic fields. \textbf{a}, The spectrum at 28~T with the single-magnon $M_0^u$, the repulsively bound two-magnon $D_\pi$ and $D_{\pi/2}$ modes (see also Fig.~\ref{fig:exp_results_2}). \textbf{b}, Zoom-in of the high-frequency spectral range corresponding to the $D_{\pi/2}$ and the repulsively bound three-magnon $T_{\pi/2}$ modes (as indicated by the arrows) measured at various fields.}
\label{fig:rawdata}
\end{figure}


\end{document}